\documentstyle[pra,aps,psfig]{revtex}
\begin{document}
\preprint{}
\draft
\title{Spectral properties of quantum $N$-body systems \\ {\em versus} \\
chaotic properties of their mean field approximations}
\author{Patrizia Castiglione,\cite{AC}
Giovanni Jona-Lasinio,\cite{JL} and Carlo Presilla\cite{AP}}
\address{Dipartimento di Fisica, Universit\`a di Roma ``La Sapienza,'' \\
Piazzale A. Moro 2, 00185 Roma, Italy}
\date{cond-mat/9604084 submitted to J. Phys. A }
\maketitle
\begin{abstract}
We present numerical evidence that in a system of interacting bosons
there exists a correspondence between the spectral properties of the 
exact quantum Hamiltonian and the dynamical chaos of the associated 
mean field evolution.
This correspondence, analogous to the usual quantum-classical correspondence,
is related to the formal parallel between the second quantization 
of the mean field, which generates the exact dynamics of the quantum
$N$-body system, and the first quantization of classical canonical coordinates.
The limit of infinite density and the thermodynamic limit are then
briefly discussed.
\end{abstract}

\section{Introduction}

The commonly accepted definition of {\sl quantum chaos} is based on
universal statistical properties of suitably defined fluctuations
in the energy spectrum. 
For instance, a confined quantum system with a finite number of degrees 
of freedom, e.g., a particle in a billiard, is said to be chaotic whenever
the nearest neighbor level spacing (NNLS) distribution, 
with the spacings normalized to their local average value, 
is well approximated by the corresponding Wigner distribution for 
random matrices \cite{GUTZWILLER}.

This definition of quantum chaos was conjectured \cite{BGS} to satisfy 
a correspondence principle with the {\sl dynamical chaos} of the associated 
classical system, i.e., the exponential sensitivity of the classical 
trajectories to a variation of the initial conditions.
In fact, a large collection of numerical examples \cite{BOHIGAS} shows 
that, whenever the classical system has positive (non positive) maximum
Lyapunov exponent, the corresponding quantum system has Wigner-like 
(non Wigner-like) NNLS distribution.
The recent result \cite{AASA} gives a theoretical support to this conjecture.

In the case of systems made of $N$ identical particles nonlinearly 
interacting, quantum chaos, in the sense stated above, seems a general rule
\cite{MKZ,BELLISS,BERK}.
For these systems, we suggest to look for a correspondence of the quantum  
chaos with the dynamical chaos of associated $c$-number canonical
coordinates well distinguished from the classical ones.
These $c$-number canonical coordinates are appropriate combinations
of time-dependent mean fields which approximate the dynamics
of the $N$-body symmetrized or antisymmetrized wavefunction \cite{BR}.
The dynamical equations of the mean fields are, in general, 
nonlinear and allow the presence of dynamical chaos as in classical 
mechanics \cite{JPC}.
A {\sl second quantization} transforms the mean fields into field operators 
and restores the exact dynamics of the quantum $N$-body system.
This second quantization can be made formally identical to the quantization
of classical canonical coordinates
and, by analogy, we can
expect a correspondence between quantum chaos of $N$-body systems 
and dynamical chaos of their mean field approximations.  

In this paper, we numerically investigate a system of $N$ bosons with both 
periodic and Dirichlet boundary conditions.
In Section II, we analyze the exact system from the point of view
of quantum chaos and find that a Wigner-like NNLS distribution is 
a general feature in presence of nonlinear interaction already for very
low values of $N$.
In Section III, we study time-dependent mean field approximations of the same
system from the point of view of the exponential sensitivity to the initial 
conditions.
We find that the mean fields show dynamical chaos in 
correspondence to the quantum chaos of the exact system
for all the values of $N$ considered.
We naturally assume that this correspondence continues to hold as
$N$ increases.
In Section IV we quantize the mean field and in 
in the last Section we discuss the limit of infinite density
and the thermodynamic limit.

\section{The model and its properties}

Let us consider a system of $N$ spinless bosons of charge $q$ 
moving in a one-dimensional lattice with $L$ sites
and described by the Hamiltonian
\begin{equation}
\hat H = \sum_{j=1}^L \left [ \alpha_j \ \hat a_j^{\dagger} \hat a_j - 
\beta_j \left( e^{i \theta} \ \hat a_{j+1}^{\dagger} \hat a_{j}
            + e^{-i \theta} \ \hat a_{j}^{\dagger} \hat a_{j+1} \right) \right] 
+ \sum_{j=1}^L  \ \gamma_j \ \hat a_{j}^{\dagger} \hat a_{j}^{\dagger}
\hat a_{j} \hat a_{j},
\label{H}
\end{equation}
where index correspondence $j \pm L=j$ is assumed.
The operator $\hat a_j^{\dagger}$ creates a boson in the site $j$ 
and $\alpha_j$, $\beta_j$, and $\gamma_j$ are 
the site, hopping, and interaction energies, respectively.
Periodic and Dirichlet boundary conditions will be considered. 
In the first case, the system represents a ring 
threaded by a line of magnetic flux $\phi$ and the phase factors are
$\theta=2 \pi \phi / \phi_0 L$, where $\phi_0=hc/q$ is the flux quantum 
(in Gauss electromagnetic system).
In the second case, the sites lie on a segment and we put 
$\beta_L=0$ and $\theta=0$.
The system (\ref{H}) has wide interest.
Its time-dependent mean-field approximation has applications to  
molecular dynamics and nonlinear optics \cite{EILBECK}
and to electron transport in heterostructures \cite{PJC}.

All properties of the system (\ref{H}) can be evaluated by knowing
eigenvalues and eigenvectors of $\hat H$.
It is simple to work in the space spanned by the Fock states 
$| n_1^i \cdots n_L^i \rangle$, where $n_j^i$ is the number of bosons
in the site $j$ and $\sum_{j=1}^L n_j^i=N$.
The index $i$ runs from 1 to the Fock dimension 
\begin{equation}
D = \frac{(N+L-1)!}{N! \ (L-1)!} 
\label{FOCKDIM}
\end{equation}
obtained by counting all the possible arrangements of the $N$ 
identical bosons into the $L$ sites.
The $D$-dimensional matrix $H$ representing the Hamiltonian (\ref{H}) 
in the Fock basis has matrix elements 
\begin{eqnarray}
H_{ki} &=& \langle n_1^k \cdots n_L^k |\hat H| n_1^i \cdots n_L^i \rangle 
= \sum_{j=1}^L \left[ \alpha_j n_j^i +
\gamma_j  n_j^i \left( n_j^i - 1 \right) \right] \delta_{ki} 
\nonumber \\ && +
\sum_{j=1}^L \beta_j \left[ 
e^{i \theta} \sqrt{n_j^i \left( n_{j+1}^i + 1 \right)} \Delta_{ki}(j) +
e^{-i \theta} \sqrt{n_{j+1}^i \left( n_j^i + 1 \right)} \Delta_{ik}(j) 
\right],
\label{HKI}
\end{eqnarray}
where
\begin{equation}
\Delta_{ki}(j) = \left\{  
\begin{array}{ll}
1  &~~ \mbox{if $n_j^k=n_j^i+1$, $n_{j+1}^k=n_{j+1}^i-1$, and 
$n_l^k=n_l^i$ for $l \neq j,j+1$}, \\
0  &~~ \mbox{otherwise}.
\end{array} \right. 
\end{equation}
The eigenvalues $E_i$ and eigenvectors $| E_i \rangle$ of the Hermitian 
matrix (\ref{HKI}) can be numerically evaluated with standard methods 
\cite{NUMREC}.
A bound to the maximum dimension $D$ that can be studied        
is essentially fixed only by the computer memory necessary to
storage the full sparse matrix (\ref{HKI}).

For a general, asymmetric system, the NNLS distribution is evaluated 
from all the eigenvalues $E_i$. 
The normalized spacings between nearest neighbor levels, 
whose distribution $P(s)$ is of interest, are taken as
\begin{equation}
s_i = \left( E_{i+1} - E_i \right) / \Delta E_{\text{av}}(i),
\label{NORMSPA}
\end{equation}
where 
\begin{equation}
\Delta E_{\text{av}}(i) = {1 \over  2 N_{\text{av}} + 1 }
\sum_{k=-N_{\text{av}}}^{N_{\text{av}}} \left( E_{i+k+1} - E_{i+k} \right)
\end{equation}
with $1 \ll N_{\text{av}} \ll D$.
In the case of Dirichlet boundary conditions, the Hamiltonian matrix 
is real symmetric and correspondence to the Gaussian orthogonal ensemble
(GOE) of random matrices may be expected.
In the case of periodic boundary conditions, the Hamiltonian matrix
is complex Hermitian and, in general, correspondence to the Gaussian 
unitary ensemble (GUE) may be expected.
However, if the external potential, represented by
the site energies $\alpha_j$, is symmetric under a reflection with 
respect to some diameter of the ring, the system has an anti-unitary 
symmetry and GOE behavior is restored \cite{RB}.

In presence of geometrical symmetries, the level statistics analysis
probing the phenomenon of level repulsion is meaningful only 
once the trivial crossings of eigenvalues belonging to different 
symmetry classes are avoided \cite{HAAKE}.
This amounts to analyze the NNLS distribution separately inside each 
one of the diagonal blocks which compose the matrix $H$ in a proper basis. 
For example, a uniform system, i.e., a system with 
$\alpha_j$, $\beta_j$, and $\gamma_j$ independent of the site-index $j$,
is invariant under rotation 
\begin{equation}
\hat {\cal R} \,\, : j \mapsto j+1 
\end{equation}
in the case of periodic boundary conditions, and space-inversion 
\begin{equation}
\hat {\cal P} \,\, : j \mapsto L+1-j 
\end{equation}
in the case of Dirichlet boundary conditions. 
The eigenvalues of $\hat H$ must be divided into $L$ classes, 
corresponding to the eigenvalues 
$\exp(i2 \pi \nu /L)$, with $\nu=1, \cdots, L$, 
of $\hat {\cal R}$, in the first case, and into two classes, 
corresponding to the 
eigenvalues $\pm 1$ of $\hat {\cal P}$, in the second one.
This is accomplished by evaluating the operator $\hat H$ 
in the basis of the degenerate eigenvectors of $\hat {\cal R}$ or 
$\hat {\cal P}$ in which the corresponding matrix $H$ is block-diagonal. 
Each block is then independently diagonalised to find the eigenvalues 
of $H$ within the corresponding symmetry class. 

The eigenvectors $| \nu_q \rangle$ of $\hat {\cal R}$, $q$ being the 
degeneracy index, are obtained by numerically solving the eigenvalue problem
\begin{equation}
\sum_{i=1}^D \left( {\cal R}_{ki} - \lambda_\nu \delta_{ki} \right)
\langle n_1^i \cdots n_L^i | \nu_q \rangle = 0
\end{equation}
where
\begin{equation}
{\cal R}_{ki} = 
\langle n_1^k \cdots n_L^k |\hat {\cal R}| n_1^i \cdots n_L^i \rangle =
\left\{  
\begin{array}{ll}
1  &~~ \mbox{if $n_j^k=n_{j+1}^i$ for $j=1,\cdots,L$}, \\
0  &~~ \mbox{otherwise}.
\end{array} \right. 
\end{equation}
The block of $H$ corresponding to the eigenvalue 
$\lambda_\nu=\exp(i2 \pi \nu /L)$ has matrix elements 
\begin{equation}
H_{\nu_q \nu_p} = \sum_{k,i=1}^D 
\langle \nu_q | n_1^k \cdots n_L^k \rangle  H_{ki}
\langle n_1^i \cdots n_L^i | \nu_p \rangle. 
\end{equation}
A similar, general procedure could be applied also to $\hat {\cal P}$.
However, the eigenvectors of $\hat {\cal P}$ are, by inspection, 
single Fock states or symmetric and antisymmetric combinations 
of couples of Fock states.
The even and odd blocks of $H$, corresponding to the eigenvalues $\pm 1$
of $\hat {\cal P}$, have matrix elements which are straightforward 
combinations of those in (\ref{HKI}).

Figure 1 shows the NNLS distribution obtained for a uniform
system with periodic and Dirichlet boundary conditions.
The distributions of the normalized spacings (\ref{NORMSPA}) 
evaluated for each symmetry class of eigenvalues have been
summed up for increasing the statistical confidence.
The agreement of the calculated NNLS distribution with the 
Wigner surmise for the GOE distribution 
$P_{\mbox{\scriptsize GOE}}(s) = (\pi s/2) \exp(- \pi s^2/4)$, 
also shown in the same figure, is statistically reliable.

The importance of performing the level statistics analysis within
the appropriate symmetry classes is evidenced in Fig. 2, 
where the NNLS distribution for the same uniform systems of Fig. 1 
is evaluated from the total spectrum of $\hat H$.
In the case of Dirichlet boundary conditions,
the mixing of the even- and odd-parity eigenvalues generates a
two-peaks distribution which behaves like $\exp(-s)$ at large $s$. 
In the case of periodic boundary conditions,
due to the higher number of symmetry classes and their statistical  
independence, we observe a distribution which mimics 
the Poisson distribution $P_{\mbox{\scriptsize P}}(s) = \exp(- s)$
even at small $s$ \cite{HAAKE}.
This fact and the observation that similar results are obtained
when the symmetries are only approximate \cite{MKZ},
will be relevant in the following.

The above mentioned geometrical symmetries can be explicitly broken 
by choosing an appropriate $j$-dependence in the parameters  $\alpha_j$, 
$\beta_j$, and $\gamma_j$.
Figure 3 shows the NNLS distribution obtained with periodic boundary 
conditions when the site energies have the form
$\alpha_j=2(N-1) \eta~ \xi_j $, where $\xi_j$ are arbitrary 
positive numbers with $\sum_{j=1}^L \xi_j =1$, and all the other parameters
are as in the uniform case. 
The resemblance of the calculated distribution with the 
GUE Wigner surmise, 
$P_{\mbox{\scriptsize GUE}}(s) = (32 s^2/\pi^2) \exp(- 4 s^2/\pi)$, 
expected on the base of the complex Hermitian nature of $H$,
is statistically reliable.
However, the GOE behavior is restored, as shown in Fig. 4, 
by choosing $\xi_{j}$ symmetric with respect to an arbitrary $j_0$.
Figures 5 and 6 show the NNLS distribution obtained 
when the hopping and interaction energies have the form indicated in 
the captions and the other parameters are as in the uniform case. 
The calculated distribution is always close to the GOE one
for both periodic and Dirichlet boundary conditions. 

Results similar to those discussed above are obtained for other choices 
of the parameters, $N$, $L$, $\alpha_j$, $\beta_j$, $\gamma_j$, and $\phi$
i.e., the system (\ref{H}) is generally characterized by a Wigner-like 
NNLS distribution.
Integrability points are the only exception to this rule.
A first, trivial, point of integrability of the system (\ref{H})
is the noninteracting case obtained for $\gamma_j \to 0$ 
with eigenvalues given by
$E_i = \sum_{j=1}^L \epsilon_j n^i_j$, 
$\epsilon_j$ being the $L$ eigenvalues obtained for $N=1$.
A second point is approached for $\beta_j \to 0$
with eigenvalues given by 
$E_i = \sum_{j=1}^L \alpha_j n^i_j  + \gamma_j n^i_j (n^i_j -1)$.
A third point of integrability is obtained by taking the continuum limit 
in which Eq. (\ref{H}) becomes the second quantization version of
the Hamiltonian of $N$ bosons interacting via a $\delta$-function potential 
\begin{equation}
\sum_{n=1}^N \left[ -{\hbar^2 \over 2 m} {\partial^2 \over \partial x_n^2}
+ C \sum_{n'=n+1}^N  \delta \left( x_n-x_{n'} \right) \right]
\label{CONTINUUM}
\end{equation}
with $0 \leq x_n \leq \ell$.
This system is solvable by the Bethe ansatz \cite{LIEB} and can be 
obtained from (\ref{H}) by putting 
$\alpha_j = 2 \beta_j = \hbar^2 / (m \Delta x^2)$ and
$\gamma_j = C/\Delta x$ with $\Delta x = \ell /L$ and letting 
$L \to \infty$.
When approaching an integrability point, the NNLS distribution transforms 
into a non Wigner-like distribution whose shape strongly depends 
on the values of the system parameters.

The results of the level statistics analysis obtained for the boson 
system (\ref{H}) are in agreement with those found in 
\cite{MKZ,BELLISS,BERK} for fermion systems.
Quantum chaos, in the sense stated in the Introduction, is a generic 
feature of systems with many particles nonlinearly interacting.
The only apparent exception to this rule is a result of
\cite{BERK} in which a system of spinless electrons moving in a ring 
similar to ours is shown to have Poisson-like NNLS distribution 
whenever the interaction is limited to a small region of the ring.
However, studying the same fermion system we found that the case
considered in \cite{BERK} has an approximate symmetry \cite{CAS}. 
When this symmetry is taken into account or is removed by changing
the relevant parameters, e.g., adding a random energy to the sites,
the NNLS distribution turns to a GOE distribution as in the example 
of Fig. 6.

\section{Mean field approximation}

Two substantially equivalent time-dependent mean field approximations 
of the Hamiltonian (\ref{H}) are obtained by choosing the $N$ particles 
to be described by the normalized boson condensate
\begin{equation}
|Z_N(t) \rangle = (N!)^{-1/2} \left[ \hat a_z^\dagger(t) \right]^N 
~|\mbox{\bf 0} \rangle 
\label{BC}
\end{equation}
or the normalized coherent state 
\begin{equation}
|Z_N(t) \rangle = \exp \left[ \sqrt{N} \hat a_z^\dagger(t) 
- \sqrt{N} \hat a_z(t)\right]
~|\mbox{\bf 0} \rangle.
\label{CS}
\end{equation}
Here, $|\mbox{\bf 0}\rangle$ is the vacuum state and $a_z^\dagger(t)$ 
creates a boson in the single-particle state 
\begin{equation}
|z(t) \rangle = a_z^\dagger(t) |\mbox{\bf 0} \rangle = 
\sum_{j=1}^L z_j(t) a_j^\dagger |\mbox{\bf 0} \rangle = 
\sum_{j=1}^L z_j(t) |j \rangle . 
\end{equation}
The normalization condition of this state, 
$\langle z(t) | z(t) \rangle = \sum_{j=1}^L |z_j(t)|^2 = 1$,
fixes the expectation number of particles in the boson condensate (\ref{BC}) 
or the coherent state (\ref{CS}) to $N$ since in both cases we have
$\langle Z_N(t) | \hat a_j^\dagger \hat a_j | Z_N(t) \rangle = N|z_j(t)|^2$.
The time evolution of the complex amplitudes $z_j(t)= \langle j|z(t) \rangle$
is obtained from a variational principle for the Dirac action \cite{BR}
\begin{equation}
\int dt~ 
\langle Z_N(t)| i \hbar {d \over dt} - \hat H |Z_N(t) \rangle. 
\label{ACTION}
\end{equation}
For a general Hamiltonian $\hat H = \hat T + \hat V$, 
sum of a single-particle term $\hat T$ and a two-particle term $\hat V$,
\begin{equation}
\hat H = \sum_{kn}  T_{kn} \hat a_k^{\dagger} \hat a_n
+\frac{1}{2}\sum_{kk'nn'} V_{kk'nn'} 
\hat a_k^{\dagger} \hat a_{k'}^{\dagger} \hat a_n \hat a_{n'}, 
\end{equation}
we have
\begin{eqnarray}
\langle Z_N(t)| i \hbar {d \over dt} - \hat H |Z_N(t) \rangle &=&
i \hbar \sum_k N~ \overline{z}_k(t) {d \over dt} z_k(t)
- \sum_{kn}  T_{kn} N~ \overline{z}_k(t) z_n(t)  
\nonumber \\ && 
- \frac{1}{2}\sum_{kk'nn'} V_{kk'nn'} 
N (N-1)~ \overline{z}_k(t) \overline{z}_{k'}(t) z_n(t) z_{n'}(t).
\end{eqnarray}
In the above expression as well as in the following ones of this Section,
we use the boson condensate (\ref{BC}).
Similar expressions hold for the coherent state (\ref{CS}) with the 
substitution $(N-1) \to N$.  
The action (\ref{ACTION}) is stationary with respect to a variation
of $\overline{z}_j(t)$ if
\begin{equation}
i \hbar {d \over dt} z_j(t) = \sum_{l} h_{jl}[z(t)] z_l(t)
\label{EZ}
\end{equation}
where
\begin{equation} 
h_{jl}[z(t)] = T_{jl} + (N-1) \sum_{kn} V_{jkln} 
~\overline{z}_k(t) z_n(t)
\end{equation}
are the matrix elements of the mean field single-particle Hamiltonian 
$h[z(t)]$.
In the case of the Hamiltonian (\ref{H}), we have
\begin{equation}
T_{jl} = \alpha_j \ \delta_{jl} 
- \beta_l e^{i\theta} \ \delta_{j l+1}  
- \beta_j e^{-i\theta} \ \delta_{j+1 l}, 
\end{equation}
\begin{equation}
V_{jkln} = 2 \gamma_j \ \delta_{jk} \ \delta_{kl} \ \delta_{ln}
\end{equation}
and the mean field single-particle Hamiltonian has matrix elements
\begin{equation}
h_{jl}[z(t)] =  \alpha_j \delta_{jl}
- \beta_{l} e^{i\theta} \delta_{j,l+1}
- \beta_{j} e^{-i\theta} \delta_{j+1,l}
+ 2 (N-1) \gamma_j |z_j(t)|^2 \delta_{jl}
\label{HJL}
\end{equation}
with $j,l=1, \ldots L$.

The system of nonlinear Schr\"odinger equations (\ref{EZ}) 
with $h_{jl}[z(t)]$ given by (\ref{HJL}) must be numerically solved 
starting from initial values $z_j(0)$ with the normalization 
condition $\sum_{j=1}^L |z_j(0)|^2 =1$.  
The choice of the numerical algorithm is critically related to
the existence of conservation laws.
By inspection, the system (\ref{EZ}) has two conserved quantities,
the single-particle probability
\begin{equation}
\| z(t) \|^2 \equiv \sum_{j=1}^L |z_j(t)|^2  
\label{CONSN}
\end{equation}
and the single-particle energy
\begin{equation}
{\cal E}[z,\overline{z}] = \sum_{j=1}^L \Bigl\{
\alpha_j |z_j(t)|^2   
- \left[ \beta_{j-1} e^{i\theta} z_{j-1}(t)
+ \beta_j e^{-i\theta} z_{j+1}(t) \right] \overline{z}_j(t) 
+ (N-1) \gamma_j |z_j(t)|^4 \Bigr\}. 
\label{CONSE}
\end{equation}
The conservation law (\ref{CONSN}) suggests the use a finite-difference
Crank-Nicholson scheme \cite{NUMREC}
\begin{equation}
\sum_{l=1}^L \left( \delta_{jl} + 
i {\Delta t \over 2 \hbar} h_{jl}[z(t+\Delta t)] \right) z_l(t+\Delta t) =
\sum_{l=1}^L \left( \delta_{jl} -
i {\Delta t \over 2 \hbar} h_{jl}[z(t)] \right) z_l(t)
\label{CRNI}
\end{equation}
which is simple to handle numerically due to the tridiagonal nature
of the matrix $h$ \cite{TRIDIAG}.
The above scheme would be correct ${\cal O}(\Delta t^2)$ 
if the matrix $h$ were time independent.
However, this is not our case and we have to approximate the matrix 
elements in the l.h.s. of (\ref{CRNI}).
The simple approximation $h_{jl}[z(t+\Delta t)] \simeq h_{jl}[z(t)]$
has catastrophic effects for the conservation law (\ref{CONSE}) 
unless very small values of $\Delta t$ are chosen.
An improvement is obtained with an iterative procedure.
Let us suppose $z_j(t)$ known and try
\begin{equation}
z_j(t+\Delta t) = \lim_{n \to \infty} z_j^{(n)}(t+\Delta t).
\end{equation}
The $n=0$ term is defined by solving 
\begin{equation}
\sum_{l=1}^L \left( \delta_{jl} + 
i {\Delta t \over 2 \hbar} h_{jl}[z(t)] \right)
z_l^{(0)}(t+\Delta t) =
\sum_{l=1}^L \left( \delta_{jl} -
i {\Delta t \over 2 \hbar} h_{jl}[z(t)] \right) z_l(t)
\label{Z0}
\end{equation}
and the $n \geq 1$ terms are chosen as solutions of
\begin{equation}
\sum_{l=1}^L \left( \delta_{jl} + 
i {\Delta t \over 2 \hbar} h_{jl}[z^{(n-1)}(t+\Delta t)] \right) 
z_l^{(n)}(t+\Delta t) =
\sum_{l=1}^L \left( \delta_{jl} -
i {\Delta t \over 2 \hbar} h_{jl}[z(t)] \right) z_l(t).
\label{ZN}
\end{equation}
This iterative scheme converges in very few steps and ensures
an excellent conservation of both quantities (\ref{CONSN}) and 
(\ref{CONSE}).

The system (\ref{EZ}) can show local exponential instability. 
The corresponding maximum Lyapunov exponent $\lambda$ defined by 
\begin{equation}
\lambda= \lim_{t \to \infty} \Lambda(t),  \ \ \ \ \ 
\Lambda(t) \equiv {1\over t} 
\ln { \| \delta z(t) \| \over \| \delta z(0) \| },
\label{LYAP}
\end{equation}
measures the exponential separation between two states 
$|z(t) \rangle$ and $|z(t) \rangle + \epsilon |\delta z(t) \rangle$ 
infinitesimally close ($\epsilon \to 0$).
The site projections $\delta z_j(t) = \langle j| \delta z(t) \rangle$ 
satisfy the set of equations obtained by linearizing (\ref{EZ})
\begin{equation}
i \hbar {d \over dt} \delta z_j(t) = \sum_{l=1}^L 
h_{jl}[z(t)] \delta z_l(t) + P \left[ z_j(t),\delta z_j(t),
\delta \overline{z}_j(t) \right],
\label{LEZ}
\end{equation}
where
\begin{equation}
P \left[ z_j(t),\delta z_j(t),\delta \overline{z}_j(t) \right] =
2 (N-1) \gamma_j z_j(t) \left[ 
z_j(t) \delta \overline{z}_j(t) + \overline{z}_j(t) \delta z_j(t) 
\right], 
\end{equation}
and can be evaluated by numerically integrating (\ref{LEZ}) 
simultaneously with (\ref{EZ}). 
The initial values $\delta z_j(0)$ can not be chosen arbitrarily.
Indeed, the state $|z(t) \rangle + \epsilon |\delta z(t) \rangle$ 
must be normalized up to terms ${\cal O}(\epsilon^2)$.
Since $\|z(t)\|=1$, we must have
$\mbox{Re} \langle z(t) | \delta z(t) \rangle =0$ at any time.
However, by using (\protect{\ref{EZ}}) and (\protect{\ref{LEZ}}) we have 
\begin{equation}
{d \over dt} \mbox{Re} \langle z(t) | \delta z(t) \rangle =
\mbox{Re} \sum_{j=1}^L \left[ 
\delta z_j(t) {d \over dt} \overline{z}_j(t) +
\overline{z}_j(t) {d \over dt}\delta z_j(t) \right] = 0
\end{equation}
and, therefore, it is sufficient to have
$\mbox{Re} \langle z(0) | \delta z(0) \rangle =0$.

For solving (\ref{LEZ}) simultaneously with (\ref{EZ})
we again adopt an iterative modification of the Crank-Nicholson scheme 
in which $P$ is considered as a driving term.
Let us suppose $z_j(t)$ and $\delta z_j(t)$ known and try
\begin{equation}
\delta z_j(t+\Delta t) = \lim_{n \to \infty} \delta z_j^{(n)}(t+\Delta t).
\end{equation}
The $n=0$ term is defined by solving 
\begin{eqnarray}
&&\sum_{l=1}^L \left( \delta_{jl} + 
i {\Delta t \over 2 \hbar} h_{jl}[z(t)] \right)
\delta z_l^{(0)}(t+\Delta t) =
\sum_{l=1}^L \left( \delta_{jl} -
i {\Delta t \over 2 \hbar} h_{jl}[z(t)] \right) \delta z_l(t)
\nonumber \\ &&~~~~~~~~~~~~~~~~~~~~~~
- i {\Delta t \over 2 \hbar}  \Big(
P \left[ z_j(t),\delta z_j(t),\delta \overline{z}_j(t) \right] +
P \left[ z_j(t),\delta z_j(t),\delta \overline{z}_j(t) \right] \Big)
\label{DZ0}
\end{eqnarray}
and the $n \geq 1$ terms are chosen as solution of
\begin{eqnarray}
&&\sum_{l=1}^L \left( \delta_{jl} +
i {\Delta t \over 2 \hbar} h_{jl}[z^{(n-1)}(t)] \right) 
\delta z_l^{(n)}(t+\Delta t) =
\sum_{l=1}^L \left( \delta_{jl} -
i {\Delta t \over 2 \hbar} h_{jl}[z(t)] \right) \delta z_l(t)
\nonumber \\ &&
- i {\Delta t \over 2 \hbar}  \left(
P \left[ z_j^{(n-1)}(t+\Delta t),\delta z_j^{(n-1)}(t+\Delta t),
\delta \overline{z}_j^{(n-1)}(t+\Delta t) \right] +
P[z_j(t),\delta z_j(t),\delta \overline{z}_j(t)] \right).
\label{DZN}
\end{eqnarray}
Note that (\ref{DZN}) must be solved after (\ref{ZN}) and (\ref{DZN})
have been solved at the iteration $n-1$ in order to know both 
$z_j^{(n-1)}(t+\Delta t)$ and $\delta z_j^{(n-1)}(t+\Delta t)$.

The quantity $\| \delta z(t) \|$ can grow exponentially and,
therefore, $\delta z_j(t)$ must be periodically scaled 
in order to avoid numerical overflows \cite{BG}. 
The scaling factors are stored for computing the Lyapunov exponent
(\ref{LYAP}).
When the system (\ref{EZ}) is chaotic, the computer round-off
errors inevitably make the numerical solutions obtained with different
integration steps $\Delta t$ different after a sufficiently 
long time.
Therefore, the comparison of solutions relative to different
steps is not a good check that the algorithm correctly works,
unless time-averaged quantities, e.g., the Lyapunov exponent (\ref{LYAP}), 
are compared \cite{BCGGS}. 
On the other hand, a check based on the conservation of the quantities 
(\ref{CONSN}) and (\ref{CONSE}) is meaningful and can be used to fix 
the size of the integration step in relation to a chosen accuracy.
With the modified Crank-Nicholson scheme described above,
for $\Delta t \lesssim 10^{-3} \hbar/\eta$
after $10^8$ iterations we have relative errors in (\ref{CONSN}) and 
(\ref{CONSE}) which are smaller than $10^{-5}$ and $10^{-4}$, respectively.

Figures 7 and 8 show the behavior of $\Lambda(t)$ with periodic
and Dirichlet boundary conditions, respectively. 
The curves denoted with $\text{u}$, $\alpha$, $\beta$, and $\gamma$,
refer to the system in which all the parameters are independent
of $j$ as in Fig. 1, 
$\alpha_j$ is random as in Fig. 3,  $\beta_j$ is as in Fig. 5, 
and $\gamma_j$ is localized as in Fig. 6, respectively.
After an initial transient, not shown in Figs. 7 and 8, $\Lambda(t)$ 
approximately stabilizes around a positive value which we take as the 
corresponding maximum Lyapunov exponent $\lambda$. 
Note that $\lambda^{-1} \lesssim 10^2 \hbar/\eta$ is much smaller 
than the maximum simulation time $10^5 \hbar/\eta$.
The value of $\lambda$ is independent of changes in the initial conditions 
$\delta z_j(0)$.
It is also independent of changes in the initial conditions $z_j(0)$ 
provided that the conserved energy ${\cal E}[z,\overline{z}]$ is not changed. 
The maximum Lyapunov exponent exceptionally vanishes when the
initial state $|z(0) \rangle$ is coincident or very close to one of 
the stationary states of the system (\ref{EZ}) which are defined by 
\begin{equation}
| z_E (t) \rangle = e^{ - {i \over \hbar} E t }
| z_E (0) \rangle,   \ \ \ \ \ \ 
\| z_E (t) \|^2 = 1.
\end{equation}

The comparison of Figs. 1-6 with Figs. 7-8 suggests a
correspondence between quantum chaos of a system of interacting
particles and dynamical chaos of its mean field approximations. 
Whenever the exact system (\ref{H}) shows Wigner-like NNLS distribution
the corresponding mean field system (\ref{EZ}), 
or that for the coherent state (\ref{CS}), has a positive maximum 
Lyapunov exponent.

The simultaneous presence of chaotic behavior in the exact and mean-field 
systems is obtained also for values of the parameters $N$, $L$, $\alpha_j$, 
$\beta_j$, $\gamma_j$, and $\phi$ different from those reported.
In the three cases in which the exact system has been shown to be integrable, 
namely $\gamma_j \to 0$, $\beta_j \to 0$, and the continuum limit, 
the corresponding mean field system is integrable and has $\lambda=0$.
For $\gamma_j \to 0$, the system (\ref{EZ}) becomes linear and 
$\Lambda(t) \equiv 0$.
For $\beta_j \to 0$, we have 
\begin{equation}
{d \over dt} |z_j(t)|^2 = z_j(t) {d \over dt} \overline{z}_j(t) +
\overline{z}_j(t) {d \over dt} z_j(t) = 0
\end{equation}
and, therefore, 
\begin{equation}
z_j(t)=z_j(0) e^{-i(\alpha_j+2(N-1)\gamma_j |z_j(0)|^2)t/\hbar}.
\end{equation}
The corresponding variation 
\begin{eqnarray}
\delta z_j(t) &=& \delta z_j(0) 
e^{-i(\alpha_j+2(N-1)\gamma_j |z_j(0)|^2)t/\hbar}
\nonumber \\ &&
- z_j(0) i {t \over \hbar} 2(N-1) \gamma_j \left[ 
\delta z_j(0) \overline{z}_j(0) + z_j(0) \delta \overline{z}_j(0)
\right] e^{-i(\alpha_j+2(N-1)\gamma_j |z_j(0)|^2)t/\hbar}
\end{eqnarray}
shows that $\| \delta z(t) \|$ is ${\cal O}(t)$ and therefore 
$\lambda=0$.
Finally, in the continuum limit the system (\ref{EZ}) becomes the
well known nonlinear Schr\"odinger equation, 
\begin{equation}
i \hbar {\partial \over \partial t} z(x,t) = -{\hbar^2 \over 2 m} 
{\partial^2 \over \partial x^2} z(x,t) + C (N-1) |z(x,t)|^2 z(x,t),
\end{equation}
solvable via spectral transform \cite{CALODEGA}.

\section{Second quantization of the mean field}
The correspondence between quantum chaos of an exact $N$-body system
and dynamical chaos of the associated mean field approximations
parallels the correspondence between quantum chaos of a single-particle
system and dynamical chaos of the associated classical equations.
This parallel can be connected to the fact 
that the first quantization of the classical canonical coordinates 
has a formal counterpart in the second quantization of the mean field 
\cite{BR}.
Let us see this in detail.
The mean field $Z_j(t) = \sqrt{N} z_j(t)$, $j=1,\ldots,L$, is determined 
by the dynamical equation
\begin{eqnarray}
i \hbar {d \over dt} Z_j(t) =
\alpha_j Z_j(t) - \beta_{j-1} e^{i\theta} Z_{j-1}(t)
- \beta_{j} e^{-i\theta} Z_{j+1}(t) 
+ 2 \gamma_j \overline{Z}_j(t) Z_j(t) Z_j(t). 
\label{NSE}
\end{eqnarray}
In this Section we will consider the coherent state (\ref{CS})  
but similar results hold for the boson condensate (\ref{BC}) with
the substitution $N \to (N-1)$.
The quantization rule
\begin{equation}
Z_j(t) \to \hat Z_j(t), ~~~~~
\overline{Z}_j(t) \to \hat Z_j^\dagger(t),
\end{equation}
with 
\begin{equation}
\left[ \hat Z_j(t) , \hat Z_k(t) \right] = 
\left[ \hat Z_j^\dagger(t) , \hat Z_k^\dagger(t) \right] = 0, ~~~~~
\left[ \hat Z_j(t) , \hat Z_k^\dagger(t) \right] = \delta_{jk} 
\label{QUANT1}
\end{equation}
transforms the nonlinear Schr\"odinger equation (\ref{NSE}) into
the Heisenberg equation for the field operator $\hat Z_j(t)$.
Indeed, in the representation of the site-localized states
$\phi_j^k = \delta_{jk}$, we have
$\hat Z_j(t) = \sum_{k=1}^L \phi_j^k \hat a_k(t) = \hat a_j(t)$ 
whose Heisenberg equation of motion is
\begin{eqnarray}
i \hbar {d \over dt} \hat a_j(t) &=& \left[ \hat a_j(t), \hat H(t) \right]
\nonumber \\ &=&
\alpha_j \ \hat a_j(t) - \beta_{j-1} \ e^{i\theta} \ \hat a_{j-1}(t)
- \beta_{j} \ e^{-i\theta} \ \hat a_{j+1}(t) 
+ 2 \gamma_j \ \hat a_j^{\dagger}(t) \hat a_j(t) \hat a_j(t). 
\end{eqnarray}

The second quantization (\ref{QUANT1}) can be made formally identical
to the first quantization of classical canonical coordinates
with the standard transformation
\begin{equation}
Q_j(t) = \sqrt{\hbar \over 2} \Big( Z_j(t) + \overline{Z}_j(t) \Big),
~~~~~P_j(t) = {1 \over i} 
\sqrt{\hbar \over 2} \Big( Z_j(t) - \overline{Z}_j(t) \Big). 
\end{equation}
By using the total energy of the system 
\begin{eqnarray}
{\cal H}[Q,P] &=& N {\cal E} [z(Q,P),\overline{z}(Q,P)] 
\nonumber \\ &=& \sum_{j=1}^L \left\{
\alpha_j {Q_j(t)^2+P_j(t)^2 \over 2 \hbar}
- \left[ \beta_{j-1} e^{i\theta}  
{Q_{j-1}(t)+iP_{j-1}(t) \over \sqrt{2 \hbar}} \right. \right.
\nonumber \\ && 
\left. \left. + \beta_j e^{-i\theta}
{Q_{j+1}(t)+iP_{j+1}(t) \over \sqrt{2 \hbar}} \right]
{Q_{j}(t)-iP_{j}(t) \over \sqrt{2 \hbar}}
+ \gamma_j \left( {Q_j(t)^2+P_j(t)^2 \over 2 \hbar} \right)^2 \right\},
\end{eqnarray}
the nonlinear Schr\"odinger equation (\ref{NSE}) can be rewritten 
in the Hamilton formalism
\begin{equation}
{d \over dt} Q_j(t) =   {d \over d P_j(t)} {\cal H}[Q,P],
~~~~~
{d \over dt} P_j(t) = - {d \over d Q_j(t)} {\cal H}[Q,P]
\label{HE}
\end{equation}
and the quantization (\ref{QUANT1}) is equivalent to the introduction of
Hermitian operators $\hat Q_j(t)$ and $\hat P_j(t)$ with commutation relations
\begin{equation}
\left[ \hat Q_j(t) , \hat Q_k(t) \right] = 
\left[ \hat P_j(t) , \hat P_k(t) \right] = 0, ~~~~~
\left[ \hat Q_j(t) , \hat P_k(t) \right] = i \hbar \delta_{jk}. 
\label{QUANT2}
\end{equation}

\section{The limit $N \to \infty$}

We first analyze the limit of infinite density, 
in which $N \to \infty$ with $L$ constant.

In this limit the quantum theory reduces to a $c$-number theory and
is, therefore, analogous to a classical limit even though $\hbar \neq 0$ 
\cite{YAFFE}.
For our system this is easily seen by retracing the $N$ dependence in the
equations of the previous section.
For large $N$, the system reduces to a collection of independent nonlinear
oscillators whose nonlinearity grows like $N$.
From the point of view of chaotic properties, the limit $N/L \to \infty$
is therefore trivial. 
An example where the same kind of limit gives rise to a non trivial
chaotic system can be found in \cite{BBZ}.

It is interesting to see the emergence of the limiting behavior 
$N/L \to \infty$ by comparing the cumulative density of states 
\begin{equation}
D(E) = {\text {Tr}}~ \theta(E-\hat H) = \sum_{i=1}^D \theta(E-E_i)
\label{DE}
\end{equation}
with the approximate expression obtained according to the Weyl rule
\begin{equation}
D_{\text{mf}}(E) = {1 \over (2 \pi \hbar)^L}
\int dQ_1 \cdots dQ_L \int dP_1 \cdots dP_L 
~\theta \left( E - {\cal H}[Q,P] \right)
~\delta \left( N - {\cal N}[Q,P] \right).
\label{DEMF}
\end{equation}
Note that we have a 
$\delta$-function constraint on the $Q-P$ phase space which fixes
\begin{equation}
{\cal N}[Q,P] \equiv \sum_{j=1}^L {Q_j(t)^2 + P_j(t)^2 \over 2 \hbar} = N
\label{CONSTRAINT}
\end{equation}
in agreement with (\ref{CONSN}).
Due to the presence of this constraint, the r.h.s. of (\ref{DEMF}) is a
$(2L-1)$-multiple integral which can be evaluated with Monte Carlo  
integration in the hypercube of side $2 \sqrt{2 \hbar N}$ centered
in the origin.
Figures 9-11 show, in a case with Dirichlet boundary conditions,
that when the density $\rho=N/L$ is increased
$D(E)$ is approximated by $D_{\text{mf}}(E)$ with increasing precision. 
Similar behavior is obtained with periodic boundary 
conditions and/or different values of the parameters of the system.
The total number of levels given by (\ref{DE}) and (\ref{DEMF}) 
become equal in the limit $N/L \to \infty$.
Indeed, we have 
\begin{equation}
D_{\text{mf}} \equiv \lim_{E \to \infty} D_{\text{mf}}(E) 
= {N^{L-1} \over \pi^L}
\int dx_1 \ldots dx_{2L} ~ \delta \left( 1 - \sum_{j=1}^{2L} x_j^2 \right)
= {N^{L-1} \over (L-1)!}
\label{DMF}
\end{equation}
which is the value of the Fock dimension (\ref{FOCKDIM}) for $N \gg L$.

For a single-particle system, the smooth behavior of the cumulative density 
of states is approximated by the corresponding semiclassical expression
in the limit of high energies ($\hbar \to 0$).
Analogously, in the case of an $N$-body system for $N/L$ large we can use
$D_{\text{mf}}(E)$ to approximate the smooth behavior of $D(E)$
and evaluate the normalized spacings (\ref{NORMSPA}) according to  
\begin{equation}
s_i = (E_{i+1} - E_i) {d \over dE} D_{\text{mf}}(E) \simeq
D_{\text{mf}}(E_{i+1}) - D_{\text{mf}}(E_i),
\end{equation}
as suggested in \cite{MKZ}.

Finally, let us briefly discuss the thermodynamic limit.
Unlike the limit $N/L \to \infty$, the system preserves its quantum 
features and the mean fields do not give a complete description.
This is also reflected by the behavior of the cumulative density of states.
For the system discussed here, when 
$N$ and $L$  $\to \infty$ with $N/L=\rho$ constant 
by using (\ref{DMF}) and (\ref{FOCKDIM}) we have
\begin{equation}
{\ln D_{\text{mf}} \over \ln D} \approx {1 + \ln \rho \over
(1 + \rho) \ln (1 + \rho) - \rho \ln \rho } \equiv \mu(\rho).
\end{equation}
The function $\mu(\rho)$ is smaller than unity for any finite $\rho$
and tends to unity for $\rho \to \infty$.
Therefore, $D_{\text{mf}} / D \approx D^{\mu(\rho) -1} $ vanishes
in the thermodynamic limit.

On the base of our numerical results and the considerations made in the
previous Section,
the correspondence between quantum chaos of an $N$-body system and
dynamical chaos of its mean-field approximations can be naturally
assumed to hold for $N \to \infty$.
This fact is exploited in \cite{JP} where the authors consider 
the chaotic behavior of the same system discussed here 
when the thermodynamic limit is approached.

\acknowledgments 
We are grateful to J. Bellissard and F. Cesi for 
enlightening remarks and suggestions. 
Partial support of INFN, Iniziativa Specifica RM6, is acknowledged.

\begin{figure}   
\centerline{\hbox{\psfig{figure=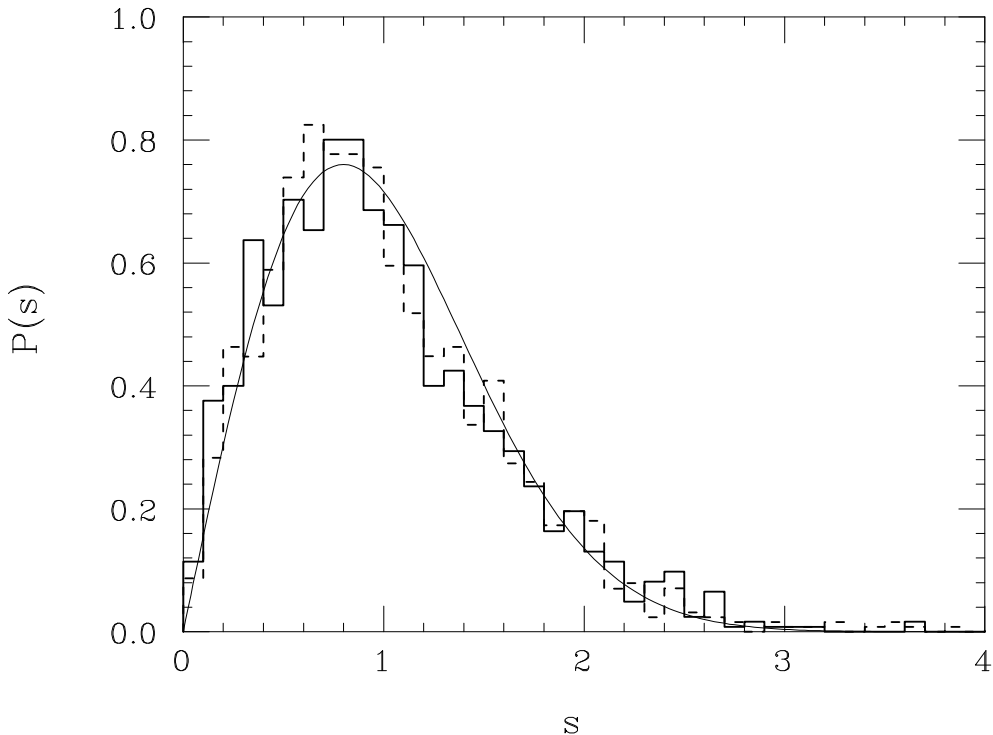,width=15.0cm,angle=90}}}
\caption{
NNLS distribution $P(s)$ for the system (\protect{\ref{H}}) in the 
uniform case 
$N=5$, $L=9$, $\alpha_j=0$, $\beta_j=\eta$, and $\gamma_j= \eta$ 
with $\phi/ \phi_0=0.3$ and periodic boundary conditions (solid histogram) 
and $\phi/ \phi_0=0$ and Dirichlet boundary conditions (dashed histogram).
The solid line is the Wigner surmise for the GOE distribution.}
\label{FIG1}
\end{figure}

\begin{figure}   
\centerline{\hbox{\psfig{figure=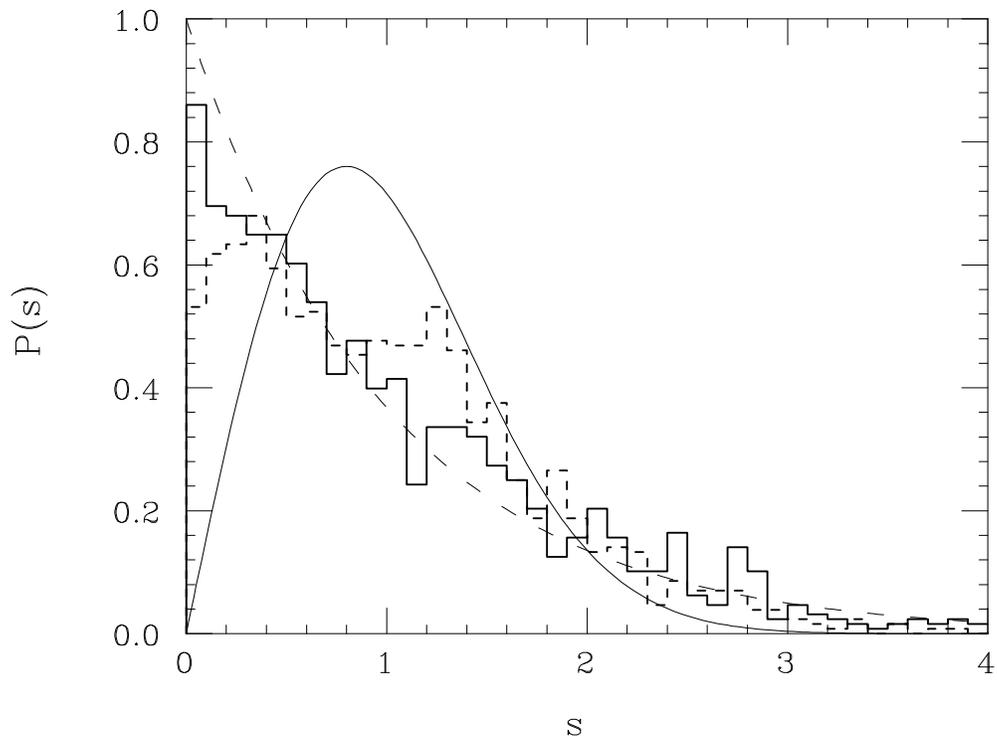,width=15.0cm,angle=90}}}
\caption{
As in Fig. 1 but without separating the eigenvalues into the appropriate
symmetry classes.
The dashed line is the Poisson distribution.}
\label{FIG2}
\end{figure}

\begin{figure}   
\centerline{\hbox{\psfig{figure=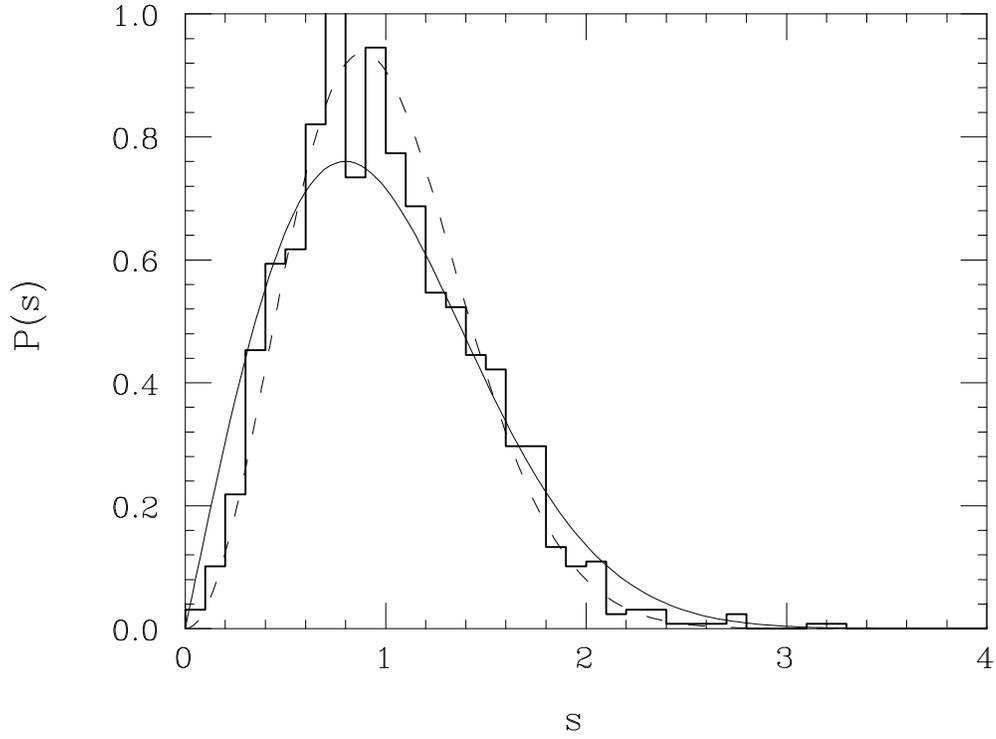,width=15.0cm,angle=90}}}
\caption{
As in Fig. 1 with periodic boundary conditions 
but choosing $\alpha_j=2(N-1) \eta~ \xi_j $, where $\xi_j$ are arbitrary 
positive numbers with $\sum_{j=1}^L \xi_j =1$.
The solid and dashed lines are the Wigner surmise for the GOE and GUE 
distributions, respectively.}
\label{FIG3}
\end{figure}

\begin{figure}   
\centerline{\hbox{\psfig{figure=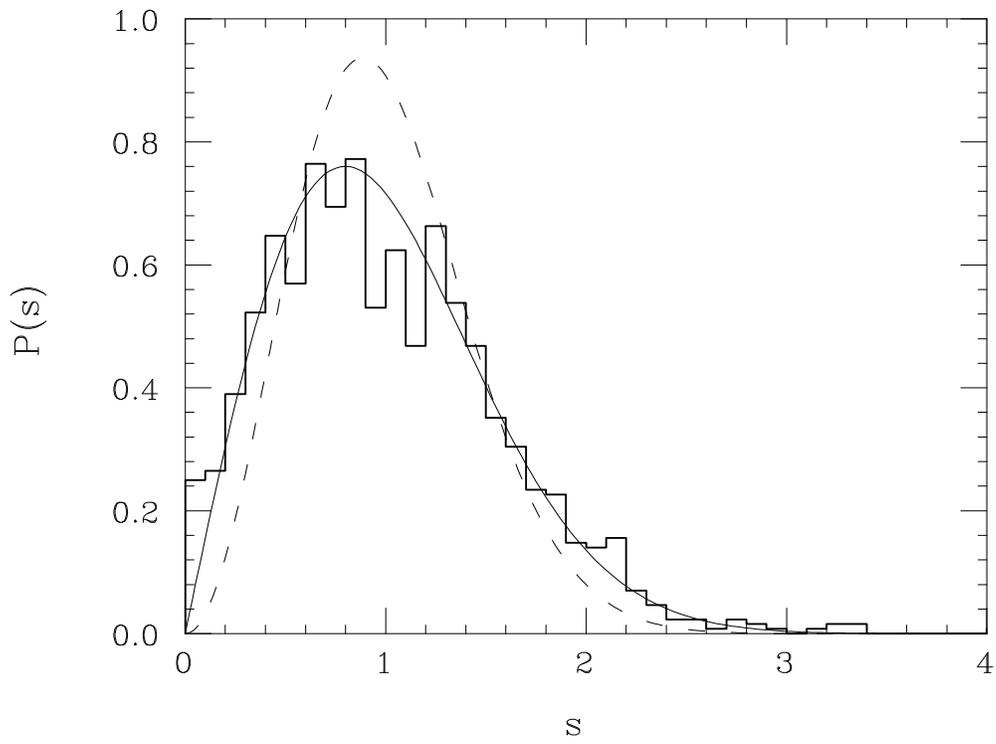,width=15.0cm,angle=90}}}
\caption{
As in Fig. 3 but choosing 
$\xi_{j_0-j}=\xi_{j_0+j}$ with $j_0$ arbitrary.}
\label{FIG4}
\end{figure}

\begin{figure}   
\centerline{\hbox{\psfig{figure=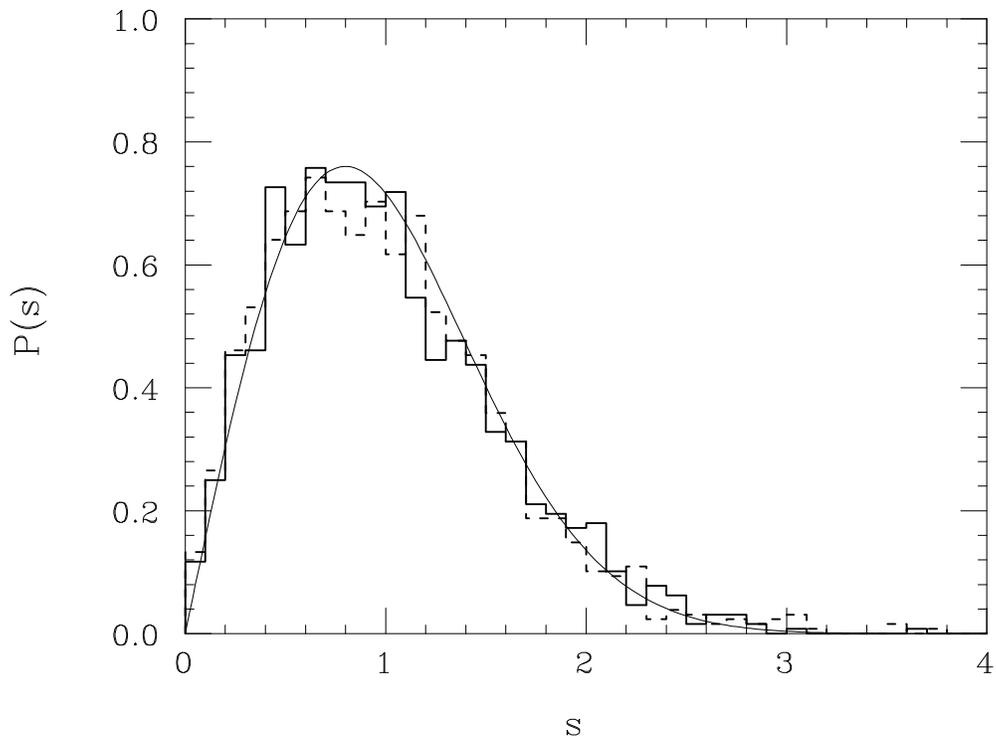,width=15.0cm,angle=90}}}
\caption{
As in Fig. 1 but choosing $\beta_2=\beta_3= 0.5 \eta$ and
$\beta_j=\eta$ for $j \neq 2,3$.}
\label{FIG5}
\end{figure}

\begin{figure}   
\centerline{\hbox{\psfig{figure=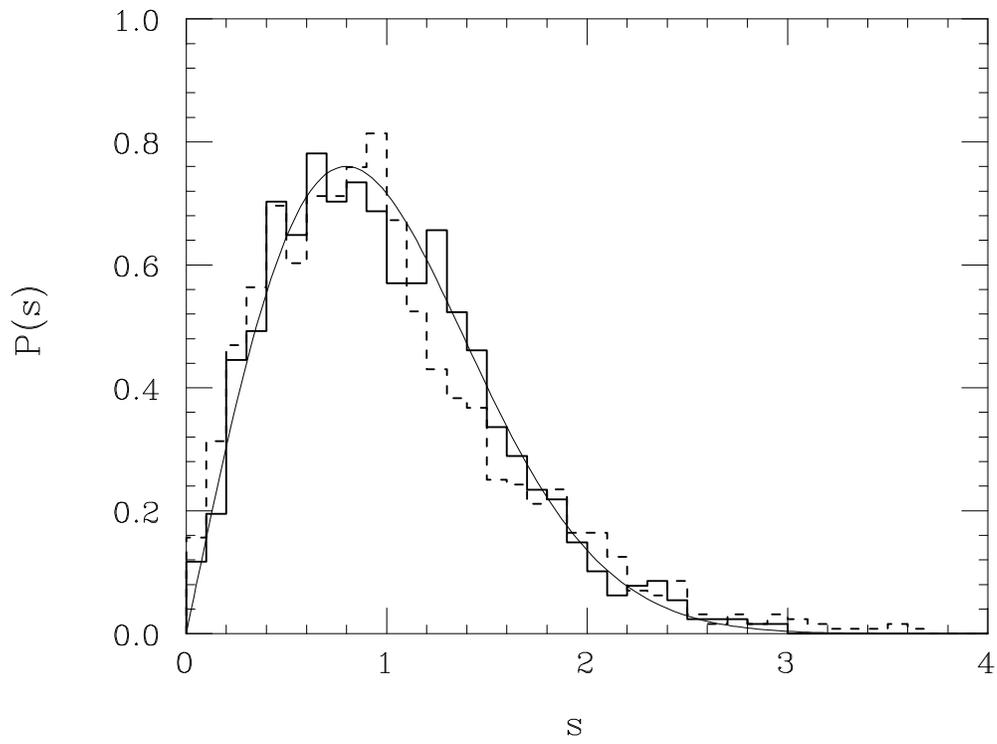,width=15.0cm,angle=90}}}
\caption{
As in Fig. 1 but choosing $\gamma_j=\eta \delta_{j3}$.}
\label{FIG6}
\end{figure}

\begin{figure}
\centerline{\hbox{\psfig{figure=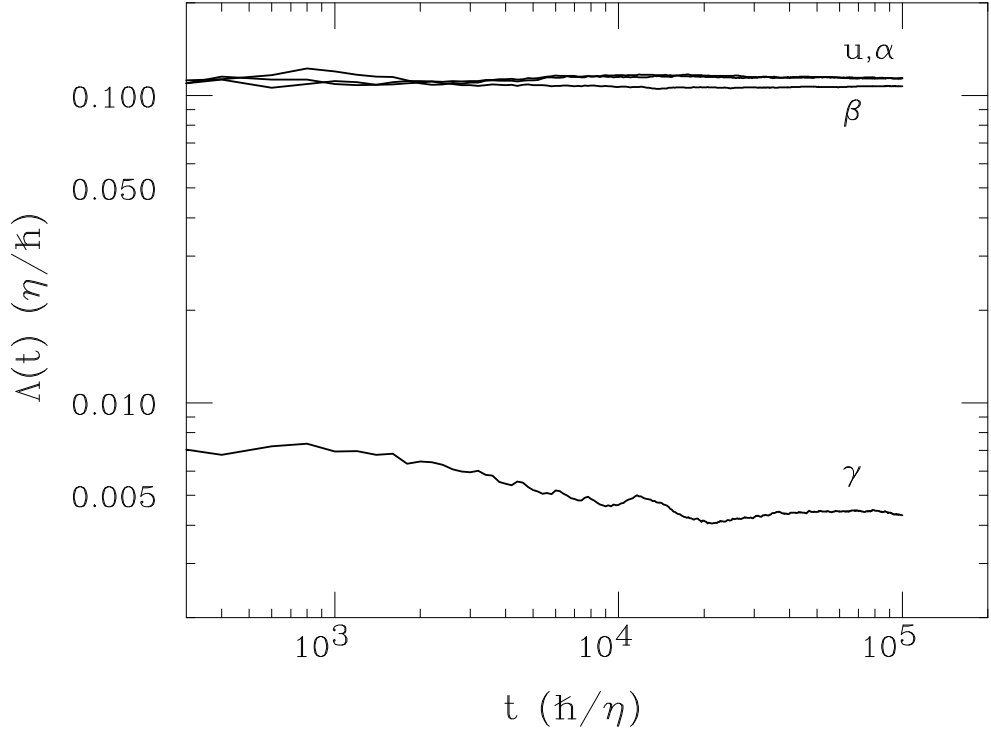,width=15.0cm,angle=90}}}
\caption{
Maximum Lyapunov exponent for the mean field system (\protect{\ref{EZ}})
with periodic boundary conditions.
The curves denoted with $\text{u}$, $\alpha$, $\beta$, and $\gamma$,
are obtained with the parameters given in Figs. 1, 3, 5, and 6, respectively.
The initial conditions $z_j(0)$ and $\delta z_j(0)$ are a set of 
arbitrary complex numbers satisfying $\| z(0) \|^2 =1$ and 
$\mbox{Re} \langle z(0) | \delta z(0) \rangle =0$.}
\label{FIG7}
\end{figure}

\begin{figure}
\centerline{\hbox{\psfig{figure=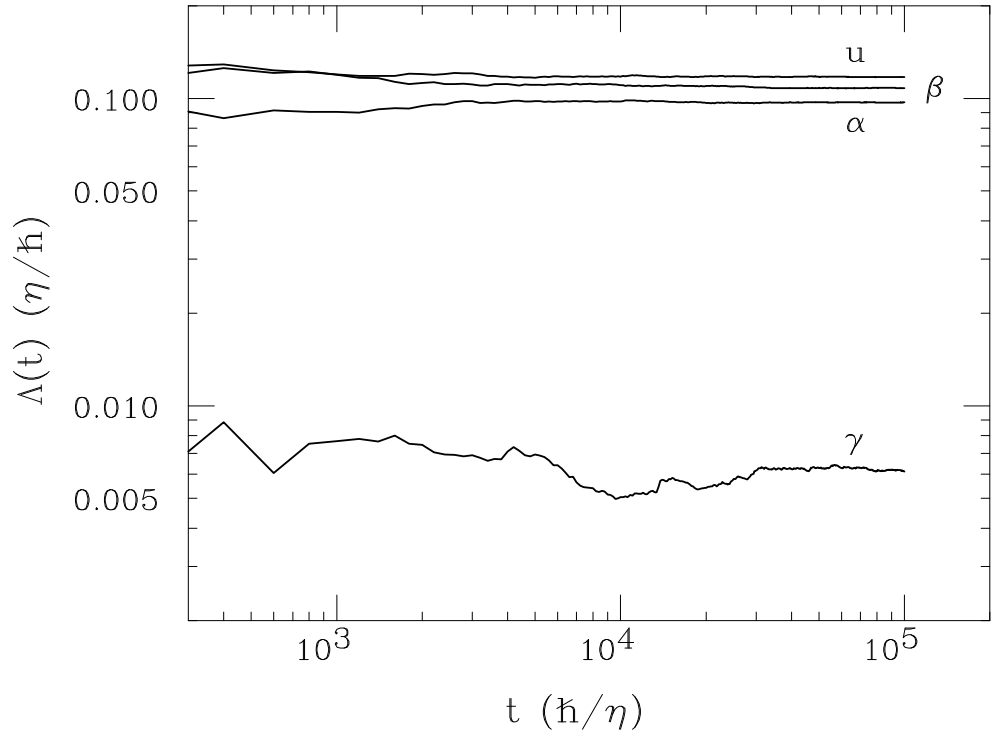,width=15.0cm,angle=90}}}
\caption{
As in Fig. 7 with Dirichlet boundary conditions.}
\label{FIG8}
\end{figure}

\begin{figure}
\centerline{\hbox{\psfig{figure=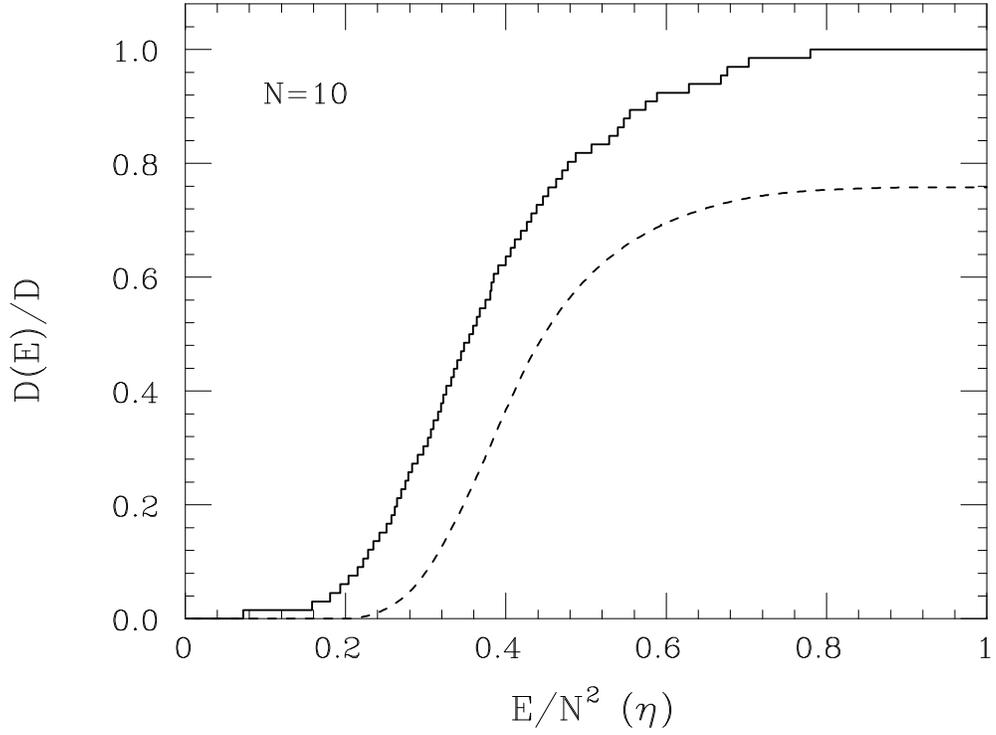,width=15.0cm,angle=90}}}
\caption{
Cumulative density of states
in the case $N=10$, $L=3$, $\alpha_j$, $\beta_j$, and $\gamma_j$
chosen between 0 and $\eta$ with a random number generator, 
and with $\phi/\phi_0=0$ and Dirichlet boundary conditions.
The histogram is the exact result (\protect{\ref{DE}})
and the dashed line is the mean field approximation (\protect{\ref{DEMF}}).}
\label{FIG9}
\end{figure}

\begin{figure}
\centerline{\hbox{\psfig{figure=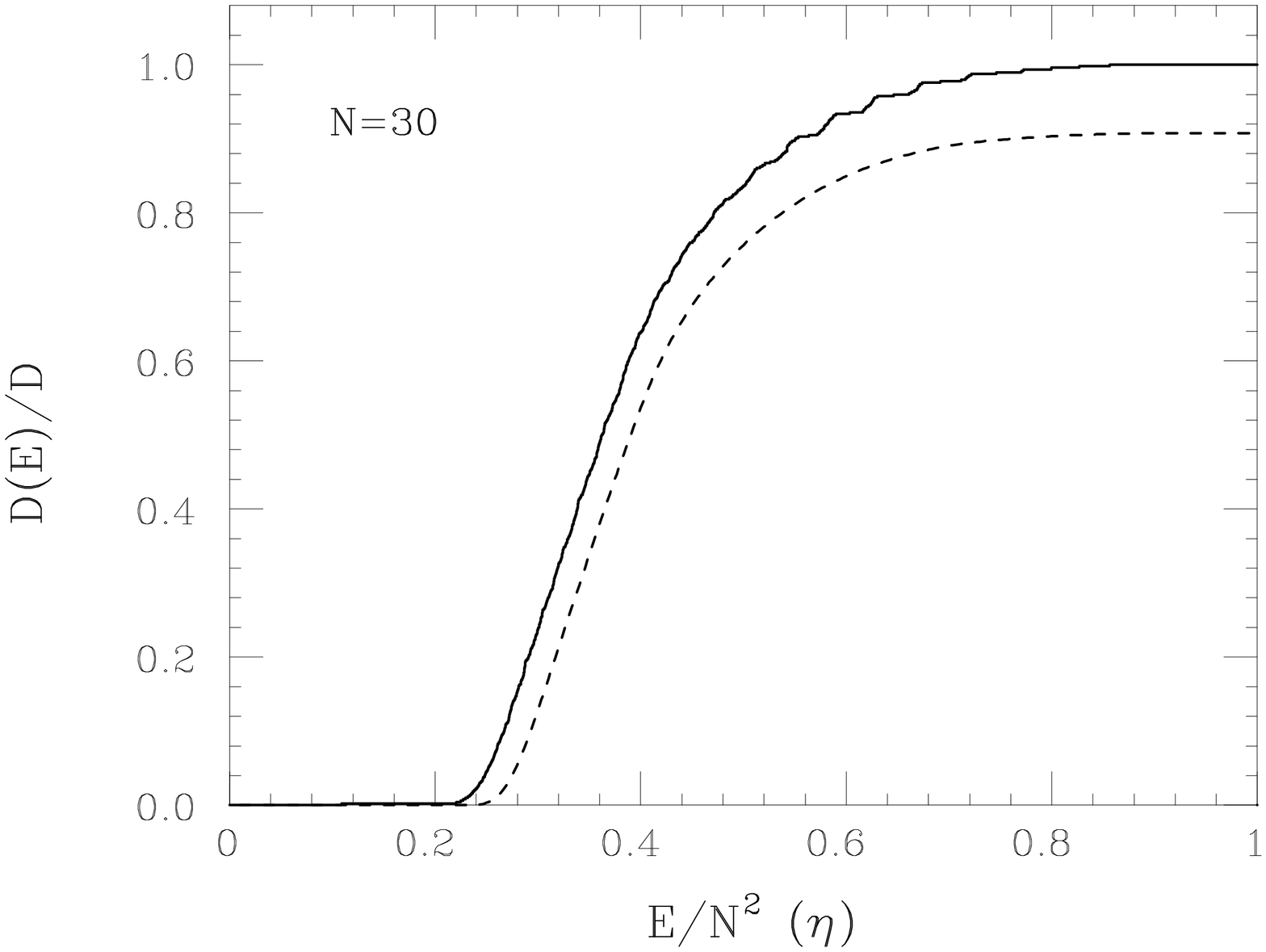,width=15.0cm,angle=90}}}
\caption{
As in Fig. 9 with $N=30$.}
\label{FIG10}
\end{figure}

\begin{figure}
\centerline{\hbox{\psfig{figure=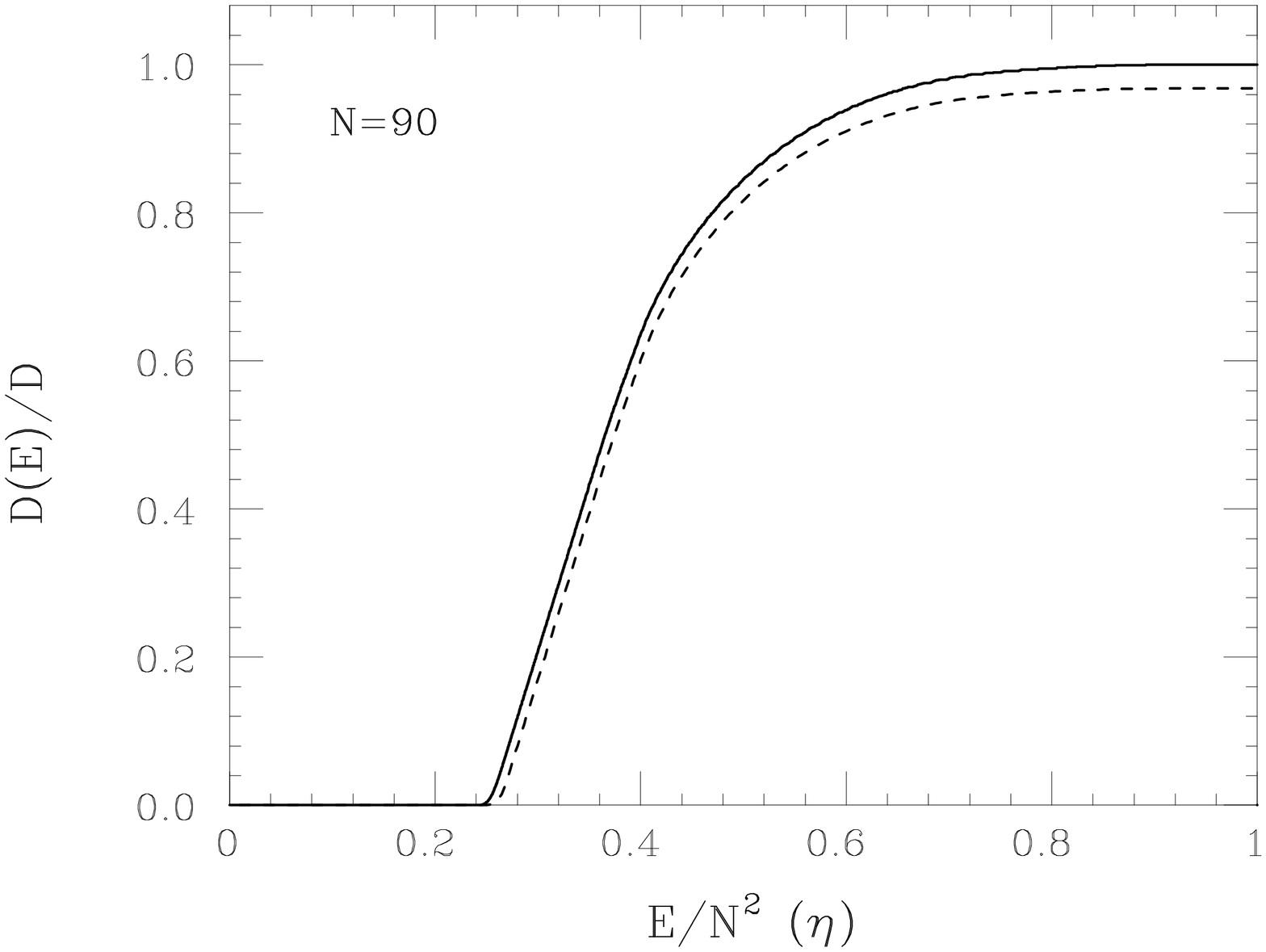,width=15.0cm,angle=90}}}
\caption{
As in Fig. 9 with $N=90$.}
\label{FIG11}
\end{figure}


\begin{references}

\bibitem[a]{AC} castiglione@roma1.infn.it

\bibitem[b]{JL} jona@roma1.infn.it

\bibitem[c]{AP} presilla@roma1.infn.it

\bibitem{GUTZWILLER} M. C. Gutzwiller, 
{\em Chaos in Classical and Quantum Mechanics}
(Springer-Verlag, Berlin, 1990).

\bibitem{BGS} O. Bohigas, M. J. Giannoni, and C. Schmit,
Phys. Rev. Lett. {\bf 52}, 1 (1984).

\bibitem{BOHIGAS} O. Bohigas in {\em Chaos and Quantum Physics},
edited by M. J. Giannoni, A. Voros, and J. Zinn-Justin,
Les Houches Summer School {\it LII}
(North-Holland, Amsterdam, 1991).

\bibitem{AASA} A. V. Andreev, O. Agam, B. D. Simons, and B. L. Altshuler,
{\em Quantum chaos, irreversible classical dynamics and random matrix 
theory}, e-print archive cond-mat/9601001.

\bibitem{MKZ} D. C. Meredith, S. E. Koonin, and M. R. Zirnbauer,
Phys. Rev. A {\bf 37}, 3499 (1988).

\bibitem{BELLISS} G. Montambaux, D. Poilblanc, J. Bellissard, and C. Sire,
Phys. Rev. Lett. {\bf 70}, 497 (1993);
D. Poilblanc, T. Ziman, J. Bellissard, F. Mila,
and G. Montambaux, Europhys. Lett. {\bf 22}, 537 (1993).

\bibitem{BERK} R. Berkovits, Europhys. Lett. {\bf 22}, 493 (1993).

\bibitem{BR} J.-P. Blaizot and G. Ripka, 
{\em Quantum Theory of Finite Systems} (MIT press, Cambridge, 1986). 

\bibitem{JPC}  G. Jona-Lasinio, C. Presilla, and F. Capasso,  
Phys. Rev. Lett. {\bf 68}, 2269 (1992).

\bibitem{EILBECK} J. C. Eilbeck, P. S. Lomdahl, and A. C. Scott,
Physica {\bf 16D}, 318 (1985);
E. Wright, J. C. Eilbeck, M. H. Hays, P. D. Miller, and A. C. Scott,
Physica D {\bf 69}, 18 (1993).

\bibitem{PJC} C. Presilla, G. Jona-Lasinio, and F. Capasso,  
Phys. Rev. B {\bf 43}, 5200 (1991).

\bibitem{NUMREC} W. H. Press, S. A. Teukolsky, W. T. Vetterling, 
and B. P. Flannery, 
{\em Numerical Recipes, the art of scientific computing} 
(Cambridge University Press, Cambridge, 1992).

\bibitem{RB} M. Robnik and M. V. Berry, J. Phys. A {\bf 19}, 669 (1986).

\bibitem{HAAKE} F. Haake, {\em Quantum Signatures of Chaos}
(Springer-Verlag, Berlin, 1992).

\bibitem{LIEB} E. H. Lieb and W. Liniger,
Phys. Rev. {\bf 130}, 1605 (1963).

\bibitem{CAS} P. Castiglione, 
Laurea Thesis, Universita di Roma ``La Sapienza,'' unpublished.

\bibitem{TRIDIAG} The matrix $h$ is tridiagonal only in the 
case of Dirichlet boundary conditions whereas in the case of periodic 
boundary conditions also the corner elements $h_{1L}$ and $h_{L1}$ 
are nonzero.
However, in this last case the lower and upper triangular (LU) 
decomposition of the matrix $h$ and the subsequent solution of 
system (\ref{CRNI}) via the forward- and back-substitution 
\protect{\cite{NUMREC}}
take only ${\cal O}(L)$ operations as in the tridiagonal case.

\bibitem{BG} G. Benettin and L. Galgani, in 
{\em Intrinsic Stochasticity in Plasmas}, 
edited by G. Laval and D. Gresillon
(Edition de Physique, Orsay, 1979).

\bibitem{BCGGS} G. Benettin, M. Casartelli, L. Galgani, A. Giorgilli,
and J.-M. Strelcyn, Nuovo Cimento {\bf 44} B, 183 (1978). 

\bibitem{CALODEGA} F. Calogero and A. Degasperis,
{\sl Spectral Transform and Solitons} 
(North-Holland, Amsterdam, 1982).

\bibitem{BBZ} G. P. Berman, E. N. Bulgakov, and G. M. Zaslavsky, 
Chaos {\bf 2}, 257 (1992).

\bibitem{YAFFE} L. G. Yaffe, Rev. Mod. Phys. {\bf 54}, 407 (1982).

\bibitem{JP} G. Jona-Lasinio and C. Presilla, 
{\em Chaotic properties of quantum many-body systems in the 
thermodynamic limit}, e-print archive cond-mat/9601056.

\end{references}
\end{document}